# Three Magnetic Rotation Bands in $^{84}$Rb

Shuifa Shen[1 2 3,*] Xin Zhao[4] Guangbing Han[5] Shuxian Wen[6] Yupeng Yan[2 3] Xiaoguang Wu[6]

Lihua Zhu[7] Chuangye He[6] Guangsheng Li[6]

([1]*Institute of Nuclear Energy Safety Technology, Chinese Academy of Sciences, Anhui, Hefei 230031*)

([2]*School of Physics, Suranaree University of Technology, Nakhon Ratchasima 30000, Thailand*)

([3]*Thailand Center of Excellence in Physics (ThEP), Commission on Higher Education, 328 Si Ayutthaya Road, Ratchathewi, Bangkok 10400, Thailand*)

([4]*Guangyuan Radiation Environmetal Monitoring Station, Sichuan, Guangyuan 628000*)

([5]*School of Physics, Shandong University, Shandong, Jinan 250100*)

([6]*China Institute of Atomic Energy, P. O. Box 275(10), Beijing 102413*)

([7]*School of Physics and Nuclear Energy Engineering, Beihang University, Beijing 100191*)

Abstract: High-spin states in $^{84}$Rb are studied by using the $^{70}$Zn($^{18}$O, p3n)$^{84}$Rb reaction at beam energy of 75 MeV. Three high-lying negative-parity bands are established, whose level spacings are very regular, i.e., there don't exist signature splitting. The dipole character of the transitions of these three bands is assigned by the γ-γ directional correlations of oriented states (DCO) intensity ratios and the multipolarity M1 is suggested by the analogy to multiparticle excitations in neighboring nuclei. The strong M1 and weak or no E2 transitions are observed. All these characteristic features show they are magnetic rotation bands.

## 1. Introduction

Magnetic rotation, a novel kind of nuclear rotations, has attracted a great interest in recent years. The levels of rotation bands are linked by strong magnetic dipole (M1) transitions whereas crossover electric quadrupole (E2) transitions are very weak. The ratios of the transition probabilities B(M1)/B(E2) are large. This magnetic character of the rotation is demonstrated by the ratios of transition probability B(M1)/B(E2) for each level in the band.

Experimental evidence of magnetic rotational bands is the presence of a greater intensity of the ΔI=1 M1 transition between neighboring levels within one band in some nuclei with small deformation. These M1 transitions differ from the ones usually observed in high-spin states. At first, the energies of these transitions are very regular, that is, there don't exist signature splitting, it is very similar to high-K rotational band in nuclei with large deformation. Secondly, their magnetic dipole reduced transition probability B(M1) values are greatly enhanced, can up to several $\mu_N^2$ units, magnetic dipole and electric quadrupole reduced transition probability ratio B(M1)/B(E2) is very large. E2 transition within the band is very weak or can not be observed, this is different from the high-K band in a well deformed nucleus, this indicates that the deformation corresponding to this band is very small, moreover, the ratio of the dynamic moment of inertia $J^{(2)}$ to electric quadrupole reduced transition probability B(E2) is larger than that of the normal deformation band and superdeformed band, it can be as large as 10 times more.

In recent years, the study of magnetic rotational band has been given great attention, either in theory or in experiments. At first, in previous years, magnetic rotational bands have been found in some nuclei in Pb region, e.g., in $^{199}$Pb and $^{200}$Pb etc., in $^{139}$Sm around A~140 mass region, in $^{110}$Cd and $^{105}$Sn around A~110 region, and in $^{82}$Rb [1] and $^{84}$Rb [2-4] around A~80 mass region etc. From the theoretical side, shell model calculations as well as relativistic mean−field (RMF) descriptions for the shears band mechanism in $^{84}$Rb were accomplished in this mass region more than ten years ago [3, 5], and its adjacent nucleus $^{82}$Rb was also studied using the *complex* Excited Vampir approach [6]. This work focuses on the magnetic dipole band in $^{84}$Rb and complements the preceding publication [7].

---

*E-mail address: shuifa.shen@fds.org.cn

## 2. Assignment of magnetic dipole band in $^{84}$Rb

High-spin states in $^{84}$Rb are studied by the heavy ion fusion-evaporation reaction $^{70}$Zn($^{18}$O, p3n)$^{84}$Rb using the $^{18}$O beam provided by the HI-13 tandem accelerator at China Institute of Atomic Energy (CIAE). Details of the experimental procedure and results were published recently in Ref. [7], where the negative-parity bands were extended greatly from the previous (6$^-$) up to the highest (17$^-$) and the spins and parities of these levels were tentatively assigned based on γ-γ directional correlations of oriented states (DCO) intensity ratios [8] and previous works. The γ-coincidence data were analyzed with the Radware software package [9]. In the following we focus on the assignment of the cascade relationship relevant to these three bands (denoted as bands C, D and E in Fig. 2 of Ref. [7]. By the way, it should be pointed out here that the γ-ray energy is 285 keV between (10$^-$) 3339 keV and (10$^-$) 3055 keV levels, 120 keV between (11$^-$) 3240 keV and 11$^{(+)}$ 3119 keV levels, and 65 keV between 11$^{(+)}$ 3119 keV and (10$^-$) 3055 keV levels, on the other hand, it should add arrow on the line to denote the 258 keV γ-ray between (14$^-$) 4698 keV and (14$^-$) 4440 keV levels, 193 keV γ-ray between (14$^-$) 4440 keV and (13$^-$) 4247 keV levels, and 345 keV γ-ray between (13$^-$) 4247 keV and (13$^-$) 3902 keV levels, they have been missed in Fig. 2 of Ref. [7] by us): In the spectrum gated by 83 keV γ-ray, besides the new found γ-rays which are used to extend positive-parity yrast band, there are still some very strong γ-rays at 185, 325, 409, 488 and 719 keV. Fig. 1 shows these γ rays clearly. Fig. 2 shows the spectrum gated by 185 keV.

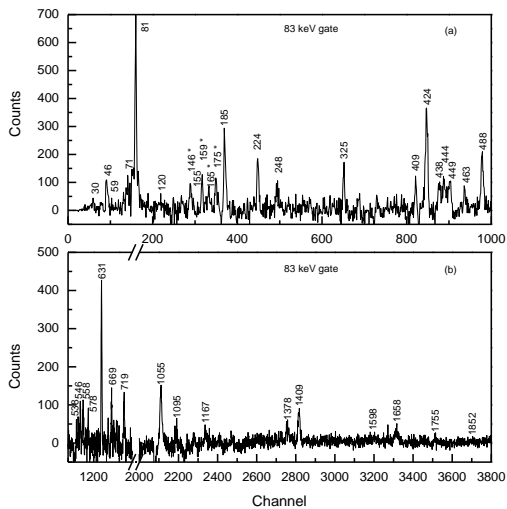

Fig. 1. Gamma-ray spectrum created by gating on the 83 keV γ-ray. (a) The low-energy region part. (b) The high-energy region. γ-rays marked with an asterisk do not belong to $^{84}$Rb.

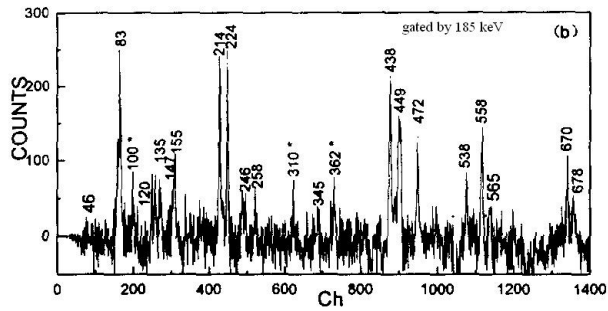

Fig. 2. Similar to Fig. 1, but for spectrum gated by 185 keV γ-ray.

One of the most caught our attention is some relatively strong γ rays at 224, 438, 449, 538, 670 keV and others. Because in the spectrum gated by 83 keV γ-ray these γ rays are also observed, furthermore the peaks are not small, so the cascade relationship between the 83 keV γ-ay and the 224, 438, 558, 449 and 538 keV γ rays has been assigned preliminarily, and also the cascade relationship between the 185 keV γ-ay and them. From the intensity relation shown in Fig. 1 it can be deduced that 185, 224 and 438 keV γ rays have direct cascade relationship. This is also proved by the spectra gated by 224 and 438 keV γ rays, at the same time, 538, 451, 345, 670 and 678 keV γ rays can be found from these two spectra, but the 558 keV γ transition can't be found, it can be seen that the 558 keV γ-ray has not the cascade relationship with them. However, from the spectrum gated by 558 keV γ-ray, 83, 46, 185 and 631 keV γ rays can be found well, and also have very strong 449, 538, 670 and 678 keV γ rays, thus states that the 558 keV γ-ray belongs to $^{84}$Rb. When gated by 449 keV γ-ray, we observe 558 keV γ-ray as expected. By comparing the spectra gated by 538, 670 and 678 keV γ-rays, we assign that the cascade relationship of 538 and 449 keV γ rays attributes to $^{78}$Kr, whereas the appearance of the 438 keV γ-ray in the spectrum gated by 449 keV γ-ray attributes to the cascade relationship of 451 and 438 keV γ rays (449 keV γ-ray is close to 451 keV γ-ray). So after careful comparison and identification, the cascade relationship is built above 185 keV γ-ray as shown in Fig. 2 of Ref. [7]. But how about the position relation between these cascades and this 83 keV γ-ray? From Fig. 2 we know that 185 keV γ-ray has no cascade relationship with 631 keV γ-ray, but has cascade relationship with 224, 558 keV γ rays and others. After the comparison of the spectra gated by 185 and 83 keV γ rays, it is found that they share the relatively high energy 1167 and 1064 keV γ rays and low energy 120 keV γ-ray, thus preliminarily assign they connect the 83 and 185 keV γ rays, then the analysis of the spectra gated by 1167, 1064 and 120 keV γ rays also proves this viewpoint. Fig. 2 also has remaining strong 214, 472, 155 keV γ lines and others. We observe 152, 472, 1275, 1295 keV γ rays and others in the spectrum gated by 214 keV γ ray, but it has no cascade relationship with 83 and 631 keV γ transitions. It can also be observed obviously the 214 and 472 keV γ transitions in the spectra gated by 224, 438, 558 keV γ transitions and others. It can thus be seen that 214 keV γ-ray is placed under the 185 keV γ-ray. Because the peak of the 472 keV γ transition is relatively good in the spectrum gated by 214 keV γ transition, so the 472 keV γ transition is placed under the 214 keV γ transition, the spectrum gated by 472 keV γ transition also proves this. Through the comparison of the spectra gated by 185, 214, 472 keV γ transitions and others, the cascade of 152, 1275, 246 and 565 keV γ transitions is built under the 214 keV γ transition, and that of 1295, 460 keV γ transitions and others under 472 keV γ transition. In addition, both 155 and 59 keV γ transitions are obviously observed in the spectra gated by 185 and 472 keV γ transitions, whereas they are not obvious in the spectrum gated by 214 keV γ transition. Because 155+59=214, so the 155 and 59 keV γ transitions are suggested to place into the level scheme as shown in Fig. 2 of Ref. [7].

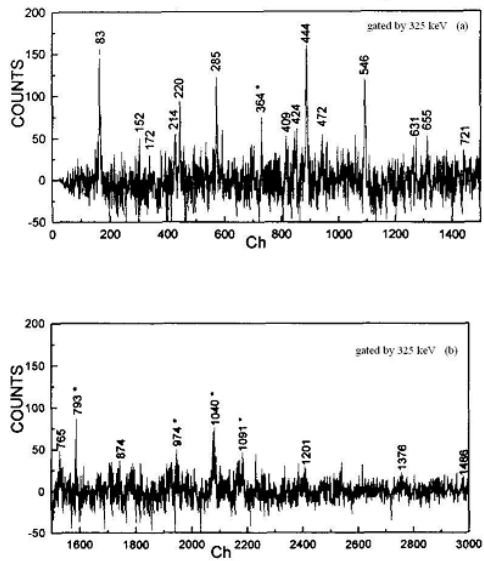

Fig. 3. Similar to Fig. 1, but for spectrum gated by 325 keV γ-ray.

Fig. 3 gives the spectrum gated by 325 keV γ transition from which very strong γ transitions at 83, 444, 546 keV and others are observed. After comparison of the spectra gated on 220, 444, 546, 655 keV γ transitions and others one by one, their transition sequence is assigned which is shown as band C in Fig. 2 of Ref. [7]. It should be noticed that in their gated spectra both 214 and 409 keV γ transitions can be observed. Except for the spectra gated by 220 and 325 keV γ transitions, 185 keV γ transition is also observed in other spectra. After comparison of the spectra gated by 325 and 214 keV γ transitions, the 285 keV γ transition is assigned between them and the 65 keV γ transition between 214 and 220 keV γ transitions. In addition in the spectrum gated by 325 keV γ transition a well 1093 keV γ transition is observed, so it is speculated that maybe a 913 keV γ transition exists between them, but in the 325 keV γ transition gated spectrum it is unobvious. Although the spectrum gated by 409 keV γ transition is more complicated, but it has relatively good 83, 631, 1376, 220, 224, 325 keV γ transitions and others, thus assign the cascade relationship among 409, 1376, 631 and 83 keV γ transitions, and the cascade relationship of 409 keV γ transition with 224, 438, 558 keV γ transitions and others through 120 keV γ transition.

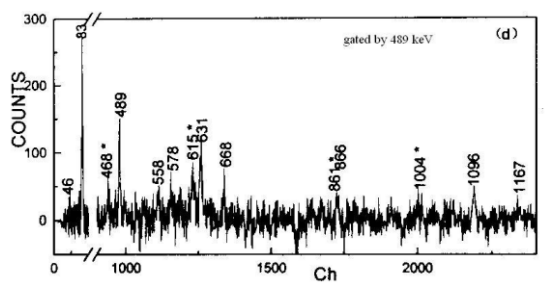

Fig. 4. Similar to Fig. 1, but for spectrum gated by 489 keV γ-ray.

There are still relatively strong 489 and 719 keV γ transitions in Fig. 1. Fig. 4 shows the spectrum gated by 489 keV γ transition, we can obviously observe the 83, 488, 631 and 1096 keV γ transitions. From the intensity it can be seen that 489 has a direct cascade relationship with 488 and 1096 keV γ transitions, respectively. The cascade relationship of 489, 1096 and 631 keV γ transitions is also can be seen in spectrum gated by 631 keV γ transition. We can also observe 1167 keV γ transition in Fig. 4, after carefully identification the 72 keV γ transition is found which is locate between 1167 and 488 keV γ transitions. In addition, there is a relatively good γ transition at 668 keV, so it can be predicted there is a 3 keV γ transition between 489 and 668 keV γ transitions. It can be proved by the observed 489 keV γ transition in both the spectra gated by 424 and 1055 keV γ transitions. The 578 keV γ transition has a relatively strong intensity in the spectrum gated by 489 keV γ transition because there exists a 248 keV γ transition between 489 and 1658 keV γ transitions. It can also be observed relatively week 224, 438, 558 keV γ transitions and others in Fig. 4, this is because 489 keV γ transition has a cascade relationship with them through 320 keV γ transition.

The remaining relatively strong transition is 719 keV γ-ray. After gating on it we find its main component belongs to $^{84}$Sr, but there are relatively good 83 and 46 keV γ transitions in its gated spectrum, this tell us that there exists 719 keV γ transition in $^{84}$Rb. After comparing with the 83 keV γ-ray gated spectrum we assign their transition sequence.

In the present work we concentrate on the most interesting feature: The sudden development of regular magnetic dipole band at excitation energy around 3 MeV (see Fig. 2 of Ref. [7]). Note that the difference about the first M1 band (denoted as Band C) between our work and the one by Schnare et al. [2-4] is one spin unit.

Among the three bands (denoted as bands C, D and E), band C is assigned as a magnetic rotational band at first. It consists of seven strong M1 transitions of 325, 444, 546, 655, 721, 765, and 874 keV on top of the $I^{\pi}=(10^-)$ level at $E_x$=3.339 MeV. The dipole character of the transitions is proven by the γ-γ directional correlations of oriented states (DCO) intensity ratios and the multipolarity M1 is suggested by the analogy to multiparticle excitations in neighboring nuclei. From its Routhian it can be found that there does not exist signature splitting. It can be observed from Fig. 1 and Fig. 3 that the 325, 444, 546, 655, 721 keV and others M1 transitions are much stronger than the 1201, 1376, 1486 keV and others E2 γ transitions. The ratios of reduced transition probabilities B(M1)/B(E2) are extracted. The error of these ratios is relatively large because the E2 transition is very weak, but the trend of the ratios can be given. The above experimental facts show that this band has a rotating magnetic characteristic. No E2 crossover transition is observed in bands D and E, which are also identified as magnetic rotational bands.

In the present work, the features of the three dipole bands shown in Fig. 2 of Ref. [7] are compared with the general criteria of a MR band in which the level energies (E) and the spins (I) follow the pattern

$$E-E_0=A_0(I-I_0)^2, \quad (1)$$

where $E_0$ and $I_0$ are the energy and spin of the band head, respectively, and $A_0$ is a constant. We plot $E-E_0$ versus $(I-I_0)^2$ in Fig. 5a, 5b and 5c for bands C, D, and E, respectively. The solid lines in the figure are the fittings of the data using the relation in Eq. (1). The good agreement of the data with the fitted curve, as shown in Fig. 5, indicates that bands C, D and E all follow the relation in Eq. (1). The dynamic moments of

the inertia $J^{(2)}$ obtained for these three bands with

$$J^{(2)}(I) \approx \frac{2\hbar^2}{\Delta E_\gamma(I)} = \frac{2\hbar^2}{E_\gamma(I+1) - E_\gamma(I)}, \qquad (2)$$

where $E_\gamma(I) \equiv E_\gamma(I \to I-1)$, are within the typical value of $J^{(2)} \sim 10\text{-}25\hbar^2\text{MeV}^{-1}$ for a MR band. Therefore, one may conclude that each of the above three bands is most likely to be an MR band.

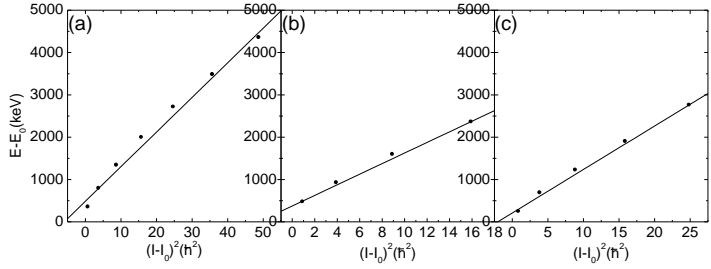

Fig. 5. Relative energy (E) versus spin (I) curve for (a) band C, (b) band D, and (c) band E, respectively. $E_0$ and $I_0$ are the bandhead energy and spin, respectively. The fitted curves are shown by the solid lines (see text for details).

## 3. Summary

High-spin states in $^{84}$Rb are populated in the reaction $^{18}$O+$^{70}$Zn at the beam energy of 75 MeV. By analyzing the γ-γ coincidence data, three negative-parity M1 sequences are observed, which show the characteristic feature of magnetic rotation such as regular level spacings and large probability ratios, of the magnetic dipole (M1) transition to the electric quadrupole (E2) transition, B(M1)/B(E2).